\documentclass[twoside]{ilcws08}
\usepackage[latin1]{inputenc}
\usepackage[dvips]{graphicx,epsfig,color}
\usepackage{wrapfig,rotating}
\usepackage{amssymb,amsmath,array}

\begin{document}
\title{ Improved Estimate of the Occupancy by Beamstrahlung Electrons
  in the ILD Vertex Detector} 
\author{Rita De Masi and Marc Winter
  \vspace{.3cm}\\
  Institut Pluridisciplinaire Hubert Curien (IPHC) \\
  23 rue du Loess - BP28- F67037 Strasbourg (France) }

\maketitle

\begin{abstract}
  The impact of the incoherent electron-positron pairs from
  beamstrahlung on the occupancy of the vertex detector (VXD) for the
  International Large Detector concept (ILD) has been studied, based
  on the standard ILD simulation tools. The occupancy was evaluated
  for two substantially different sensor technology in order to
  estimate the importance of the latter. The influence of an anti-DID
  field removing backscattered electrons has also been studied.
\end{abstract}

\section{Introduction}
The incoherent production of electron-positron pairs resulting from
the beam-beam interaction is the main source of background for the ILD
vertex detector, and it is most constraining for its innermost layer.
These electrons and positrons are produced with a longitudinal
momentum up to few hundreds GeV and a transverse momentum of few tens
of MeV on average. Due to their low $p_T$, they spiralize in the
solenoidal magnetic field, whose field lines are parallel to the beam
line, thus several of them can traverse repeatedly the same VXD layer.
Those primary electrons and positrons may also hit elements of the
detector further down the beam line, originating low energy particles
traveling backward (secondaries), which may reach the VXD. The rate of
secondaries reaching the VXD depends strongly on the presence of an
additional dipole field located further down the beam line, as shown
in Section~\ref{aD}; thus primaries and secondaries will be analized
separately in the following. The time when the hit has taken place,
will be used to distinguish them. Namely, will be considered as
generated by primaries all hits with a hit time shorter than 20~ns and
by secondaries those with a hit time larger than 20~ns.  A detailed
description of this analysis can be found in \cite{bkgnote}.

\section{Analysis}
100 bunch crossings (BX) generated with the GuineaPig \cite{GP}
generator have been studied. The standard simulation and
reconstruction tools for the ILD detector concept have been used (i.e.
Mokka \cite{Mokka} and Marlin \cite{Marlin} respectively). The model
of detector used in this study takes properly into account the angle
of 14~mrad between the beam directions. A preliminary description of
the calorimeters along the beam line is also included \cite{Mokka}.

\subsection{Hit density}
\label{sec:ht}
The number of hits in the first layer of the VXD as function of the
coordinate along the beam line $z$ and the polar angle $\phi$ is shown
in Fig.~\ref{Fig::occl1}. Besides a change of the absolute hit rate,
analogous distributions can be observed for the remaining layers. The
$\phi$ distribution shows a significant increase of the number of hits
in the region $|\phi|<50^\circ$, due to the particles with large hit
time which are not produced symmetrically around the $z$ axis.  The
"spikes" in the $\phi$ distributions are due to particle crossing the
overlapping regions of neighbouring ladders.

\subsection{Occupancy}
\begin{wrapfigure}{r}{0.5\columnwidth}
\vspace*{-0.4cm}\centerline{\includegraphics[width=0.5\columnwidth]
{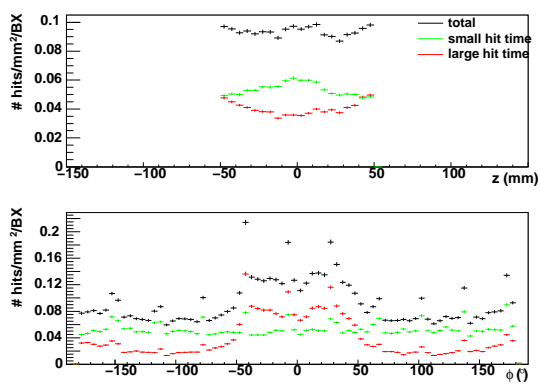}}
\vspace*{-0.4cm}
\caption{Number of hits per mm$^2$ per BX for the first layer as
      function of $z$ (upper plot) and $\phi$ (lower plot). The green
      and red plots correspond to a cut on short and large hit time
      particles respectively.}\label{Fig::occl1}
\end{wrapfigure}
In order to calculate the occupancy, the effective path length of the
particles inside the sensitive volume of the detector ougth to be
accounted for. It may reach up to several millimeters, especially for
backscattered particles which were produced at small polar angle in
order to reach the VXD. \\
The occupancy depends on the characteristics of the VXD, namely pixel
size, integration time, number of hit pixels per impact, effective
thickness of the sensitive volume. In absence of choice of the sensor
technology, a set of those parameters has been agreed upon in the ILD
vertex community as reference and they have been used to estimate the
occupancy. As a comparison, the occupancy has been also estimated in
the framework of a specific technology (CMOS~\cite{cmos}). The
parameters describing both options are shown in Tab.~\ref{Tab::sC}).
\begin{table}
\begin{tabular}{|c|c|c|c|c|}
    \hline
    layer&\multicolumn{2}{c|}{standard} &\multicolumn{2}{c|}{CMOS} \\
    \cline{2-5}

     & pitch ($\mu$m) & integration time ($\mu$s) & pitch ($\mu$m) & integration time ($\mu$s) \\
    \hline
    1 & 25 &  50 & 20 &  25\\
    2 & 25 & 200 & 25 &  50\\ 
    3 & 25 & 200 & 33 & 100\\ 
    4 & 25 & 200 & 33 & 100\\ 
    5 & 25 & 200 & 33 & 100\\ 
    \hline
  \end{tabular}
  \caption{Parameters of the VXD layers for the standard and CMOS
    configuration. 50~$\mu$m and 15~$\mu$m sensitive thickness, 3 and 
    5 hit pixels in average for straight impact respectively.}
\label{Tab::sC}
\end{table}

\noindent The results for the occupancy in each layer are shown for
the two configurations in Tab.~\ref{Tab::occupancy}. 
\begin{table}
\begin{tabular}{|c|c|c|c|c|c|c|}
    \hline
    layer&\multicolumn{3}{c|}{standard} &\multicolumn{3}{c|}{CMOS} \\
    \cline{2-7}
    & total & large hit time & short hit time & total & large hit time & short hit time\\
    \hline
    1 & 0.0790 & 0.0347 & 0.0443 & 0.0183 & 0.0080 & 0.0103 \\
    2 & 0.0381 & 0.0164 & 0.0217 & 0.0062 & 0.0026 & 0.0035 \\ 
    3 & 0.0105 & 0.0049 & 0.0056 & 0.0054 & 0.0025 & 0.0029 \\ 
    4 & 0.0041 & 0.0020 & 0.0021 & 0.0021 & 0.0010 & 0.0011 \\ 
    5 & 0.0016 & 0.0006 & 0.0010 & 0.0008 & 0.0003 & 0.0005 \\ 
    \hline
  \end{tabular}
  \caption{Occupancy for each layer in absence of an anti-DID for the 
    standard and CMOS configurations. The large and small hit time 
    components are shown, as well as their sum.}
\label{Tab::occupancy}
\end{table}
\noindent The values are averaged over $\phi$. In fact, due to the
$\phi$ dependence shown in Figure~\ref{Fig::occl1}, the local
occupancy in a $\phi$ sector can be twice as high as the mean. In
average, one can conclude that the large hit time contribution to the
occupancy is more than $40\%$ of the total rate.

\subsection{Anti-DID magnetic field}
\label{aD}
A Detector Integrated Dipole (anti-DID), aligning the outgoing beam
with the experimental magnetic field, can be used to reduce the beam
size growth due to synchrotron radiation. The anti-DID impacts also
the hit rate on the VXD due to beamstrahlung electrons, by reducing
the number of backscattered electrons travelling backwards from
further along the beam line. 

\begin{wrapfigure}{r}{0.5\columnwidth}
\vspace*{-0.cm}\centerline{\includegraphics[width=0.5\columnwidth]
{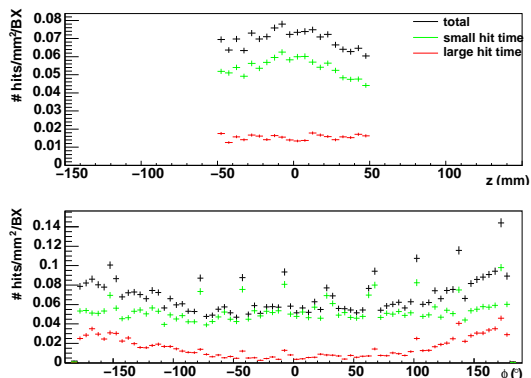}}
\vspace*{-0.5cm}
\caption{Number of hits per mm$^2$ per BX for the first layer as
  function of $z$ (upper plot) and $\phi$ (lower plot) in presence of
  an anti-DID field. }\label{Fig::occl1aD}
\end{wrapfigure}

\noindent The anti-DID reduces by roughly 30\% the number of hits on
the VXD, in particular the large hit time component, as can be seen in
Figure~\ref{Fig::occl1aD}. This leads to a more homogeneous local
distribution in $\phi$.
The occupancy of the ILD vertex detector, which is a driving
parameter of its requirements, has been evaluated with the latest
version of the experimental apparatus, assuming a five-layer VXD
geometry with 15~mm inner radius and a 3.5~T magnetic field. The
evaluation was performed for two different sets of pixel
characteristics, representative of the most mature sensor technologies
under consideration. Both sets assume a continous read-out during the
train.  They differ by their read-out time, pixel pitch, cluster
multiplicity and sensitive volume thickness.
\section{Conclusion}
Occupancies of $\sim2\%$ and $\sim7\%$ were found in the innermost
layer for the two sets. The average occupancy would be about 30\%
lower in presence of anti-DID, with a 50\% decrease in one azimuthal
sector. Accounting for the uncertainties on these predictions
translates into upper limits on the occupancy in the innermost layer
in the range 5-15\%, depending on the sensor characteristics. These
high rates plead for additional R\&D on the sensors equipping this
layer, in particular for shortening the read-out time significantly
below 50~$\mu s$.

\begin{footnotesize}

\end{footnotesize}

\end{document}